\documentclass[aps,pre,reprint,onecolumn,notitlepage,groupedaddress]{revtex4-1}
\usepackage{graphics,amssymb,amsmath,color,bm,mathtools,upgreek}
\usepackage[colorlinks=true,linkcolor=blue]{hyperref}
\usepackage[sfdefault=cmbr]{isomath}

\def \t{\tensorsym}
\def \lb{\left}
\def \rb{\right}

\def \d{\,\text{d}}

\def \etah{\hat{\eta}}

\def \bgamma{\boldsymbol{\gamma}}

\def \bgammad{\dot{\boldsymbol{\gamma}}}
\def \bgammadh{\hat{{\dot{\bgamma}}}}

\def \bnabla{\boldsymbol{\nabla}}

\def \bOmega{\mathbf{\Omega}}
\def \bOmegah{\hat{\mathbf{\Omega}}}

\def \bsigma{\boldsymbol{\sigma}}
\def \bSigma{\boldsymbol{\Sigma}}
\def \bsigmah{\hat{\boldsymbol{\sigma}}}

\def \btau{\boldsymbol{\tau}}

\def \btauh{\hat{\boldsymbol{\tau}}}

\def \bXih{\hat{\mathbf{\Xi}}}
\def \bzero{\mathbf{0}}

\def \bA{\mathbf{A}}

\def \bB{\mathbf{B}}

\def \fB{\mathcal{B}}

\def \bE{\mathbf{E}}
\def \bEh{\hat{\mathbf{E}}}

\def \tE{\mathsf{\t E}}
\def \tEh{\mathsf{\t{\hat{E}}}}

\def \bF{\mathbf{F}}
\def \bFh{\hat{\mathbf{F}}}

\def \tF{\mathsf{\t F}}
\def \tFh{\mathsf{\t{\hat{F}}}}

\def \tGh{\mathsf{\t{\hat{G}}}}

\def \bL{\mathbf{L}}
\def \bLh{\hat{\mathbf{L}}}

\def \bRh{\hat{\mathbf{R}}}

\def \tRh{\mathsf{\t{\hat{R}}}}

\def \bS{\mathbf{S}}
\def \bSh{\hat{\mathbf{S}}}

\def \tTh{\mathsf{\t{\hat{T}}}}

\def \bU{\mathbf{U}}

\def \bUh{\hat{\mathbf{U}}}

\def \tU{\mathsf{\t U}}
\def \tUh{\mathsf{\t{\hat{U}}}}

\def \fV{\mathcal{V}}

\def \bn{\mathbf{n}}

\def \br{\mathbf{r}}

\def \bu{\mathbf{u}}
\def \buh{\hat{\bu}}

\def \bx{\mathbf{x}}

\begin{document}

\title{Force moments of an active particle in a complex fluid}
\author{Gwynn J. Elfring}\email{Electronic mail: gelfring@mech.ubc.ca}
\affiliation{
Department of Mechanical Engineering, Institute of Applied Mathematics 
University of British Columbia, Vancouver, BC, V6T 1Z4, Canada}
\begin{abstract}
A generalized reciprocal theorem is formulated for the motion and hydrodynamic force moments of an active particle in an arbitrary background flow of a (weakly nonlinear) complex fluid. This formalism includes as special cases a number of previous calculations of the motion of both passive and active particles in Newtonian and non-Newtonian fluids.
\end{abstract}
\maketitle

\section{Introduction}
The reciprocal theorem for low Reynolds number hydrodynamics is a method in which one utilizes a (seemingly) unrelated `auxiliary' flow field in order to dramatically simplify calculations in Stokes flows \citep{happel65}. For example, the reciprocal theorem can be used to find the motion of passive particles in background flows, without solving for the disturbance flow the particle generates, by instead using the flow the particle generates under rigid-body motion in an otherwise quiescent fluid. This approach was extended to obtain the motion of an active particle, one that is deforming or has a surface slip velocity, in a background flow \citep{stone96}. Recently a method was developed to use the reciprocal theorem to find the first moment of the hydrodynamic force on the surface of an active particle in a quiescent fluid \citep{lauga16}. In this case the auxiliary flow is the disturbance flow generated by a passive particle held fixed in a linear background flow.

The reciprocal theorem has also been adapted to describe the motion of particles in non-Newtonian flows (see \citet{leal79, leal80} and references therein). In this case the disturbance flow of a particle in a background flow of a Newtonian fluid is solved and used to obtain the weakly nonlinear non-Newtonian result without solving the flow field in the complex fluid. This approach was extended to find the motion of active particles in otherwise quiescent fluids \citep{lauga09,lauga14} and background flows \citep{datt17} and may also be used to describe the motion of multiple particles in complex fluids \citep{elfring14} (and Newtonian fluids \citep{papavassiliou15}). Recently, the reciprocal theorem has also been used to find the first force moment on a passive sphere in a viscoelastic linear flow \citep{koch06, rallison12, koch16, einarsson17b}.

In this note, we show that all these examples may be treated as special cases under a more general framework. We use a generalized reciprocal theorem to find the motion and all force moments of an active (or passive) particle (or particles) in a non-Newtonian (or Newtonian) flow. This approach does not depend on a specific choice of auxiliary flow and renders all force moments in a transparent manner. 

\section{A generalized reciprocal theorem}
Consider a free particle $\fB$ with surface $\partial \fB$ immersed in a background flow $\bu^\infty$ (note that $\bu^\infty$ describes the background flow without the presence of the particle) of a complex fluid with deviatoric stress
\begin{align}
\btau = \eta \bgammad +\epsilon\bSigma
\end{align}
where $\eta$ is the viscosity and $\bgammad$ is the strain-rate tensor. The symmetric tensor $\bSigma[\bu]$ is a nonlinear functional of the velocity field $\bu$ while $\epsilon$ is a small dimensionless parameter indicating the magnitude of the departure from linear behavior \citep{leal80}. This constitutive relationship may be interpreted as a perturbation from a Newtonian flow field, where $\epsilon$ may be the Carreau number or viscosity ratio for shear-thinning fluids \citep{datt15} or Deborah number for weakly viscoelastic fluids. This form is natural for `weak, slow' flows of viscoelastic fluids \citep{leal80}, but more general viscoelastic constitutive relationships may be recast in this form in Fourier space \citep{lauga14} where $\eta$ would be complex. We assume that $\eta$ is constant in space but viscosity variations may be embedded in $\bSigma$. The particle may be active so that the velocity boundary condition on the particle is
\begin{align}
\bu(\bx\in \partial\fB) = \bU+\bOmega\times\br+\bu^S,
\end{align}
where $\bU$ and $\bOmega$ are the rigid body translation and rotation of the body, $\br = \bx -\bx_0$, where $\bx_0$ is a convenient point on the body (for instance, the center of mass), while $\bu^S$ is a velocity due to surface activity, such as diffusiophoretic slip \citep{michelin14} or a swimming gait (see \citet{elfring14} for a detailed description of swimming kinematics). 

Consider as an auxiliary (or dual, or complementary) problem, here denoted by a hat, a body of the same instantaneous shape moving as a rigid body, $\buh(\bx\in \partial\fB) = \bUh+\bOmegah\times\br$, in a background flow $\buh^\infty$ of a Newtonian fluid, $\btauh=\etah\bgammadh$. To simplify matters we define disturbance flows $\bu' = \bu -\bu^\infty$ and $\buh' = \buh -\buh^\infty$ and also disturbance stresses $\bsigma'=\bsigma-\bsigma^\infty$ and $\bsigmah'=\bsigmah-\bsigmah^\infty$. Note that the disturbance stress $\bsigma'$ is not necessarily equivalent to the stress field associated with $\bu'$ due to the possible nonlinearity of the constitutive equation. We consider all flow fields to be incompressible and neglect inertia in the fluid so that
\begin{align}
\bnabla\cdot\bu'=\bnabla\cdot\buh'=0,\\
\bnabla\cdot\bsigma'=\bnabla\cdot\bsigmah'=\bzero.\label{divsig}
\end{align}
From the above we can construct the following identities from the products
\begin{align}
(\bnabla\cdot\bsigma')\cdot\buh'=(\bnabla\cdot\bsigmah')\cdot\bu'=0,
\end{align}
and upon rearranging we can write equivalently
\begin{align}
\bnabla\cdot(\bsigma'\cdot\buh')-\btau':\bnabla\buh'=\bnabla\cdot(\bsigmah'\cdot\bu')-\btauh':\bnabla\bu'=0,
\end{align}
where $\bA:\bB = \sum_{ij}A_{ij}B_{ij}$. Following substitution of the constitutive relations and use of the identity $\bgammadh':\bnabla\bu'=\bgammad':\bnabla\buh'$ the two equations may be combined to obtain
\begin{align}
\bnabla\cdot\lb(\bsigma'\cdot\buh'-\frac{\eta}{\etah}(\bsigmah'\cdot\bu')\rb)=\epsilon\bSigma':\bnabla\buh'.
\end{align}

Integrating over the volume of fluid $\fV$ exterior to $\fB$ and applying the divergence theorem we obtain
\begin{align}
\int_{\partial\fV}\bn\cdot\lb(\bsigma'\cdot\buh'-\frac{\eta}{\etah}(\bsigmah'\cdot\bu')\rb)\d S=-\frac{1}{2}\int_{\fV}\epsilon\bSigma':\bgammadh' \d V,
\end{align}
where the surface $\partial \fV$ that bounds the fluid volume $\fV$ is composed of the surface of the body, $\partial \fB$ on the interior, and an outer surface. Here, $\bn$ is the normal to the surface, $\partial \fV$, pointing \textit{into} $\fV$. Provided the fields, $\bu'$ and $\bsigma'$, decay appropriately in the far-field \citep{leal80}, or the fluid is bounded by no-slip walls \citep{becker96}, we may neglect the outer surface so that $\partial\fV\rightarrow \partial \fB$ . If the fluid were Newtonian, $\epsilon\bSigma'=\bzero$, the above is then a standard version of the reciprocal theorem for two Newtonian fluids and indicates the equality of the \textit{virtual powers} of the motion of $\partial\fB$ and a dual flow field \citep{happel65}.

Applying the boundary conditions on the surface on $\partial\fB$ we obtain
\begin{align}\label{total1}
\bF'\cdot\bUh+\bL'\cdot\bOmegah-\frac{\eta}{\etah}\lb(\bFh'\cdot\bU +\bLh'\cdot\bOmega\rb)-\int_{\partial\fB}\bn\cdot\bsigma'\cdot\buh^\infty\d S
=\frac{\eta}{\etah}\int_{\partial\fB}\bn\cdot\bsigmah'\cdot\lb(\bu^S-\bu^\infty\rb)\d S-\frac{1}{2}\int_{\fV}\epsilon\bSigma':\bgammadh'\d V,
\end{align}
where the force and torque are respectively
\begin{align}
\bF' = \bF-\bF^\infty &= \int_{\partial\fB}\bn\cdot\bsigma'\d S,\\
\bL' = \bL-\bL^\infty &= \int_{\partial\fB}\br\times(\bn\cdot\bsigma')\d S.
\end{align}
Moreover, because $\bF^\infty=\bL^\infty=\bzero$, we drop the primes on these terms.

A key point is that it is useful to express the background flow of the auxiliary problem as a Taylor series in $\br=\bx-\bx_0$, 
\begin{align}\label{taylorseries}
\buh^\infty=\bUh^\infty+\br\cdot(\bEh^\infty+\bXih^\infty)+\ldots \ ,
\end{align}
where $\bEh^\infty$ and $\bXih^\infty$ are the symmetric and antisymmetric parts of the velocity gradient tensor of the background flow at $\bx_0$. In this way \eqref{total1} becomes
\begin{align}\label{total2}
\bF\cdot\bUh'+\bL\cdot\bOmegah'-\bS':\bEh^\infty-\ldots-\frac{\eta}{\etah}\lb(\bFh'\cdot\bU +\bLh'\cdot\bOmega\rb)
=\frac{\eta}{\etah}\int_{\partial\fB}\bn\cdot\bsigmah'\cdot\lb(\bu^S-\bu^\infty\rb)\d S-\frac{1}{2}\int_{\fV}\epsilon\bSigma':\bgammadh'\d V,
\end{align}
where $\bUh' = \bUh - \bUh^\infty$ and $\bOmegah' = \bOmegah - \bOmegah^\infty$ while
\begin{align}
\bS' = \bS-\bS^\infty = \int_{\partial \fB}\frac{1}{2}\lb[\bn\cdot\bsigma'\br+\br\bn\cdot\bsigma'\rb] \d S
\end{align}
is the symmetric first moment (or \textit{stresslet} \citep{batchelor70a}) of the hydrodynamic traction on the body due to the disturbance stress. The zeroth moment and the antisymmetric first moment lead to a force and torque absorbed into the first two terms of \eqref{total2}.

Introducing a more compact notation, where
\begin{align}
\tF' &=\lb[\bF, \ \bL, \ \bS', \ \ldots\rb],\\
\tFh' &=\lb[\bFh, \ \bLh, \ \bSh', \ \ldots\rb],\\
\tUh' &=\lb[\bUh', \ \bOmegah', \ {-\bEh^\infty}, \ \ldots\rb],\\
\tU  &= \lb[\bU, \ \bOmega, \ \bzero, \ \ldots \rb],
\end{align}
we obtain
\begin{align}
\tF'\cdot\tUh'-\frac{\eta}{\etah}\tFh'\cdot\tU=\frac{\eta}{\etah}\int_{\partial\fB}\bn\cdot\bsigmah'\cdot\lb(\bu^S-\bu^\infty\rb)\d S-\frac{1}{2}\int_{\fV}\epsilon\bSigma':\bgammadh'\d V.\label{totalnot1}
\end{align} 
Now, by the linearity of the flow of the Newtonian auxiliary problem, we may write
\begin{align}
\buh' &= \tGh\cdot\tUh',\label{linearG}\\
\bgammadh'/2&=\tEh\cdot\tUh',\label{linearE}\\
\bsigmah' &= \tTh\cdot\tUh',\label{linearT}\\
\tFh' &= -\tRh\cdot\tUh'\label{linearR}.
\end{align}
These are, in principle, infinitely large tensors (for a completely a general flow); $\tGh$, $\tEh$ and $\tTh$ are functions of position in space that map $\tUh'$ to the fluid velocity, strain-rate and stress fields respectively, while the the grand resistance tensor, $\tRh$, connects velocity moments to force moments, in the auxiliary problem. Substitution of these linear relationships into \eqref{totalnot1}, upon discarding the abritrary $\tUh'$, leads to our main result
\begin{align}
\frac{\eta}{\etah}\tRh\cdot\tU=-\tF'+\frac{\eta}{\etah}\int_{\partial\fB}\lb(\bu^S-\bu^\infty\rb)\cdot(\bn\cdot\tTh)\d S-\int_{\fV}\epsilon\bSigma':\tEh\d V.
\label{main}
\end{align}
Equation \eqref{main} is a tensorial relationship for the motion and \textit{all} force moments of an active (or passive) particle in an arbitrary background flow of a non-Newtonian (or Newtonian) fluid. The details of the dual problem are irrelevant in that no specific body motion nor background flow needs to be chosen, only the linear operators $\tEh$, $\tTh$ and $\tRh$ enter the picture (even the viscosity does not matter as all instances of $\etah$ cancel in \eqref{main}). Indeed this is a fundamental point, the solution cannot depend on a particular choice of auxiliary problem. A clever choice of auxiliary problem can be useful to reduce the dimensionality of the problem \textit{a priori}, but that ability is not lost here and is equivalent to a carefully chosen inner product; although, care should be taken because the rheology of the fluid can lead to non-trivial coupling between translation and rotation, even for symmetric particles \citep{oppenheimer16}. We also note that this same equation extends directly for systems of $N$ bodies, the derivation remains essentially unchanged except $\partial \fB$ instead represents the union of the surfaces of all bodies and tensors are extended to account for all bodies, for example $\tU$ would include the translation and rotation of all $N$ bodies (see \citet{elfring14} and \citet{papavassiliou15} for the multibody mobility problem). While the functional form of \eqref{main} is identical, resolution of the $N$-body Newtonian problem can be considerably more complicated.

In an alternative form we may split the volume integral, upon an application of the divergence theorem, into a boundary term and a volume term to obtain
\begin{align}
\frac{\eta}{\etah}\tRh\cdot\tU=-\tF'+\frac{\eta}{\etah}\int_{\partial\fB}\lb(\bu^S-\bu^\infty\rb)\cdot(\bn\cdot\tTh)\d S+\epsilon\tF_\Sigma'+\int_{\fV}(\bnabla\cdot\epsilon\bSigma')\cdot\tGh \d V,
\label{main2}
\end{align}
where the term $\epsilon\tF_\Sigma'$ represents moments of the non-Newtonian stress $\epsilon\bSigma'$. Although less compact, this form allows one to differentiate between the non-Newtonian stress acting directly on the body from the effects of the modification of the flow due to the complex rheology \citep{einarsson17, einarsson17b}.

Because the tensor $\bSigma$ depends on the velocity field $\bu$, non-Newtonian problems are often tackled perturbatively, $\bu = \bu_0 +\epsilon \bu_1+\ldots$. In this way, the non-Newtonian term will depend only on the Newtonian solution to leading order $\bSigma[\bu_0]$. The volume integral in \eqref{main2} may still be unwieldy but symmetry arguments or a carefully chosen basis may simplify analytical resolution (see for example \citet{einarsson17}) and numerical integration is straightforward. This formalism may also be extended to account for weak inertial effects, whereby the small parameter is then the Reynolds number \citep{cox70,leal80,becker96}, but care should be taken due to the singular nature of a weakly inertial expansion.

\section{Rigid-body motion}
In the generalized reciprocal relationship given by \eqref{main} or \eqref{main2}, the vector $\tU$ contains only rigid-body motion and because of this, only the rigid-body resistance tensor 
\begin{align}
\tRh_{\tF\tU} = 
\begin{bmatrix}
\bRh_{FU} & \bRh_{F\Omega}\\
\bRh_{LU} & \bRh_{L\Omega}
\end{bmatrix},
\end{align}
need be inverted to solve the mobility problem. In other words, we may first solve the mobility problem in isolation
\begin{align}
\tU=\frac{\etah}{\eta}\tRh_{\tF\tU}^{-1}\cdot\lb[-\tF+\frac{\eta}{\etah}\int_{\partial\fB}\lb(\bu^S-\bu^\infty\rb)\cdot(\bn\cdot\tTh)\d S-\int_{\fV}\epsilon\bSigma':\tEh\d V\rb],
\label{mobility}
\end{align}
where here $\tU$, $\tF$ are six-dimensional only (in the case of a single particle) comprising translation/rotation and hydrodynamic force/torque respectively, while $\tTh$ and $\tEh$ are similarly truncated \citep{elfring16,datt17}. If the inertia of the body is negligible (small Stokes numbers) then the hydrodynamic force must balance any external or interparticle force  $\tF = - \tF_{P}$ \citep{brady88}, and so we may alternatively write
\begin{align}
\tU=\frac{\etah}{\eta}\tRh_{\tF\tU}^{-1}\cdot\lb[\tF_P+\tF_T+\tF_{NN}\rb],
\end{align}
where
\begin{align}
\tF_T = \frac{\eta}{\etah}\int_{\partial\fB}\lb(\bu^S-\bu^\infty\rb)\cdot(\bn\cdot\tTh)\d S
\end{align}
is a Newtonian `thrust' from the surface activity and `drag' from the background flow, while
\begin{align}\label{forcenn}
\tF_{NN} = -\int_{\fV}\epsilon\bSigma':\tEh\d V = \epsilon\tF_\Sigma'+\int_{\fV}(\bnabla\cdot\epsilon\bSigma')\cdot\tGh \d V
\end{align}
represents the extra non-Newtonian force/torque on the particle.

The external force may be prescribed (for example weight due to gravity) and the motion computed, or vice versa, the motion of the particle may be prescribed and the hydrodynamic force computed. If the particle is passive, $\bu^S=\bzero$, then we have an equation for the motion of a passive particle in a complex flow \citep{leal80}. If the particle is active, $\bu^S\ne\bzero$, we recover the result of \citet{stone96} for a Newtonian fluid, $\tF_{NN}=\bzero$, and Lauga \textit{et al.} for a complex fluid \citep{lauga09,lauga14,elfring14, datt17}. In the case of two particles, \citet{khair10} showed that by prescribing the rigid-body motion of two passive spheres in an otherwise quiescent fluid and measuring of the non-Newtonian force which arises from \eqref{forcenn} one may extract normal stress coefficients for weakly nonlinear flows.

\section{Force moments}
Upon resolution of the mobility problem represented by \eqref{mobility}, the higher force moments may then be determined individually. For example, from \eqref{main} the symmetric first moment is given by
\begin{align}
\bS'= \bS-\bS^\infty = -\frac{\eta}{\etah}\lb[\bRh_{SU}\cdot\bU+\bRh_{S\Omega}\cdot\bOmega\rb]
+\frac{\eta}{\etah}\int_{\partial\fB}\lb(\bu^S-\bu^\infty\rb)\cdot(\bn\cdot\tTh_E)\d S-\int_{\fV}\epsilon\bSigma':\tEh_{E}\d V,
\label{stresslet}
\end{align}
where $\bRh_{SU}$ and $\bRh_{S\Omega}$ connect rigid-body motion with the symmetric first moment in a Newtonian fluid, whereas the subscript $E$ indicates the part of the operator associated with $-\bEh^\infty$ in \eqref{linearE} and \eqref{linearT} (the fourth-order tensor $\tE_{E}$ is contracted $-\tE_{E}:\bEh^\infty$ to give the rate of strain associated with $-\bEh^\infty$). By way of \eqref{stresslet} one may obtain the stresslet of a passive or active particle in a Newtonian or non-Newtonian fluid in an arbitrary background flow; below we demonstrate two particular cases from the recent literature. Higher-order force moments are obtained in similarly systematic fashion by considering the corresponding higher-order terms of the expansion of the background flow in \eqref{taylorseries} (for example, second moments are obtained from the quadratic term).
For a swimmer in an otherwise quiescent ($\bu^\infty=\bzero$) Newtonian fluid ($\epsilon\bSigma'=\bzero$) the symmetric first moment,
\begin{align}
\bS = -\frac{\eta}{\etah}\lb[\bRh_{SU}\cdot\bU+\bRh_{S\Omega}\cdot\bOmega\rb]+\frac{\eta}{\etah}\int_{\partial\fB}\bu^S\cdot(\bn\cdot\tTh_{E})\d S,
\end{align}
was recently derived by \citet{lauga16} by using a linear auxiliary flow. For spherical particles $\bRh_{SU}=\bRh_{S\Omega}=\bzero$ and so for squirmer-type swimmers only the right-hand side integral remains and $\tTh_{E}$ is well known for a sphere in straining flow.

For the symmetric first moment due to a passive sphere ($\bu^S=\bzero$) in a complex fluid we have instead
\begin{align}
\bS' = -\frac{\eta}{\etah}\int_{\partial\fB}\bu^\infty\cdot(\bn\cdot\tTh_{E})\d S-\int_{\fV}\epsilon\bSigma':\tEh_{E}\d V.
\end{align}
If the background flow is an unbounded linear (straining) flow, $\bu^\infty = \bE^\infty\cdot\br$ then the first term on the right-hand side yields $\frac{\eta}{\etah}\bSh'=3\eta \bE^\infty V_\fB$, where $V_\fB$ is the volume of $\fB$, while
\begin{align}
\bS^\infty =\int_\fB \bsigma^\infty\d V = 2\eta\bE^\infty V_\fB+\int_\fB\epsilon\bSigma^\infty\d V.
\end{align}
Altogether we obtain
\begin{align}
\bS &= 5\eta \bE^\infty V_\fB + \int_{\fB}\epsilon\bSigma^\infty \d V-\int_\fV \epsilon\bSigma':\tEh_{E} \d V,\nonumber\\
&= 5\eta \bE^\infty V_\fB + \int_{\fB}\epsilon\bSigma^\infty \d V+\bS_\Sigma'+\int_{\fV}(\bnabla\cdot\epsilon\bSigma')\cdot\tGh_{E}\d V
\label{stresslet2}
\end{align}
which corresponds to recent results \citep{einarsson17b}, except that here we used disturbance fields for the non-Newtonian flow to obviate the evaluation of surface integrals in the far-field. To recover the form given by \citet{einarsson17b}, we use far field asymptotic results from that work to obtain
\begin{align}
\bS^\infty_\Sigma=\int_\fB \epsilon \bSigma^\infty\d V -\int_\fV(\bnabla\cdot\bSigma^\infty)\cdot\tGh_{E}\d V,
\end{align}
which, when combined with \eqref{stresslet2}, leads to
\begin{align}
\bS &= 5\eta \bE^\infty V_\fB +\bS_\Sigma+\int_{\fV}(\bnabla\cdot\epsilon\bSigma)\cdot\tGh_{E}\d V.
\end{align}

\section{Conclusions}
In this note we have developed a generalized reciprocal theorem to find the motion and all force moments of an active (or passive) particle in a non-Newtonian (or Newtonian) fluid. This general approach eliminates the selection of specific auxiliary flows for certain cases and instead focuses on the linear operators which connect force moments with velocity moments for particles in Newtonian fluids. This approach encompasses a number of previous problems and provides a framework to attack more complex problems in the future. We note that certainly not all problems that are described by this framework have been solved. The effect of fluid rheology on a single active particle with general deformation, or slip boundary conditions is far from understood \citep{riley15, elfring15, datt17}. The motion of two active particles in quiescent Newtonian fluids was recently studied by means of the reciprocal theorem, both for diffusiophoretic particles \citep{sharifi-mood16}, and squirmers \citep{papavassiliou17}, likewise hydrodynamic interactions between two passive spheres in weakly nonlinear fluids \citep{khair10, pak12}, but hydrodynamic interactions amongst two (or many) active bodies in non-Newtonian fluids remain largely unexplored and can lead to quantitatively different dynamics \citep{elfring10}; the same can be said of active particles in non-Newtonian background flows \citep{mathijssen16}. If the particles are Brownian, fluid rheology can lead to significant differences in observed trajectories \citep{gomez-solano16} and while it is possible to include a stochastic force in the above theory, one requires proper description of that forcing in nonlinear non-Newtonian flows \citep{squires10, zia15}.

\acknowledgements
The author acknowledges funding from NSERC and thanks S\'ebastien Michelin for helpful discussions and suggestions.

\newpage

\bibliography{swimming}
\end{document}